\documentclass[letterpaper,conference]{IEEEtran}

\def\BibTeX{{\rm B\kern-.05em{\sc i\kern-.025em b}\kern-.08emT\kern-.1667em\lower.7ex\hbox{E}\kern-.125emX}}
\usepackage{import} 
\usepackage{nicefrac}
\usepackage{siunitx}
\usepackage{array,framed}
\usepackage{booktabs}
\usepackage{
  color,
  float,
  epsfig,
  wrapfig,
  graphics,
  graphicx,
  subcaption
}
\usepackage{textcomp,amssymb}
\usepackage{setspace}
\usepackage{latexsym,fancyhdr,url}
\usepackage{enumerate}
\usepackage{algorithm2e}
\usepackage{graphics}
\usepackage{xparse} 
\usepackage{xspace}
\usepackage{multirow}
\usepackage{csvsimple}
\usepackage{balance}

\usepackage{
  tikz,
  pgfplots,
  pgfplotstable
}
\usepackage{hyperref}

\usetikzlibrary{
  shapes.geometric,
  arrows,
  external,
  pgfplots.groupplots,
  matrix
}

\pgfplotsset{compat=1.9}

\usepackage{mathtools,}

\DeclareMathAlphabet{\mathcal}{OMS}{cmsy}{m}{n}

\DeclareGraphicsExtensions{%
    .png,.PNG,%
    .pdf,.PDF,%
    .jpg,.mps,.jpeg,.jbig2,.jb2,.JPG,.JPEG,.JBIG2,.JB2}

\newcommand{\EN}{{EN}\xspace}
\newcommand{\ENnospace}{{EN}}
\newcommand{\ONnospace}{{EG}}
\newcommand{\ENs}{{ENs}\xspace }
\newcommand{\ONs}{{EGs}\xspace }
\newcommand{\ON}{{EG}\xspace }
\newcommand{\eg}{{\em e.g.,}\ }
\newcommand{\ie}{{\em i.e.,}\ }
\newcommand{\sol}{{\em CLEDGE}\xspace}

\usepackage{color, xcolor}
\newcommand{\susmit}{\color{black}}
\newcommand{\spyros}{\color{black}}
\newcommand{\fgcs}{\color{black}}

\usepackage{etoolbox}
\usepackage{xstring}
\DeclareListParser{\doslashlist}{/}
\newcounter{ndnNameComponentCounter}%
\newcommand{\name}[1]{{%
  \setcounter{ndnNameComponentCounter}{0}%
  \renewcommand{\do}[1]{{%
    \ifnumgreater{\value{ndnNameComponentCounter}}{0}{\allowbreak/}{}%
    \ifnumodd{\value{ndnNameComponentCounter}}{}{}%
    ##1}%
    \stepcounter{ndnNameComponentCounter}}%
``{\fontfamily{cmtt}\small\selectfont\IfBeginWith{#1}{/}{/}{}\doslashlist{#1}}''%
}}

\makeatletter
\def\ps@IEEEtitlepagestyle{%
  \def\@oddfoot{\mycopyrightnotice}%
  \def\@evenfoot{}%
}
\def\mycopyrightnotice{%
  {\footnotesize \textcolor{red}{\begin{tabular}[t]{@{}l@{}} This paper has been accepted for publication by the 46th IEEE Conference on Local Computer Networks (LCN). © 2021 IEEE. Personal use of this material \\ is permitted. Permission from IEEE must be obtained for all other uses, in any current or future media, including reprinting/republishing this material for \\ advertising or promotional purposes, creating new collective works, for resale or redistribution to servers or lists, or reuse of any copyrighted component \\ of this work in other works.\end{tabular}}}
  \gdef\mycopyrightnotice{}
}

\IEEEpubid{\begin{tabular}[t]{@{}l@{}}zzz\\zzz\\zzz\end{tabular}}

\title{CLEDGE: A Hybrid Cloud-Edge Computing Framework over Information Centric Networking}

\author{Md Washik Al Azad,\thanks{Md Washik Al Azad is with the University of Nebraska, Omaha (email: \href{mailto:malazad@unomaha.edu}{malazad@unomaha.edu}).}
     Susmit Shannigrahi,\thanks{Susmit Shannigrahi is with the Tennessee Tech University (email: \href{mailto:sshannigrahi@tntech.edu}{sshannigrahi@tntech.edu}).}
             Nicholas Stergiou,\thanks{Nicholas Stergiou is with University of Nebraska, Omaha (email: \href{mailto:nstergiou@unomaha.edu}{nstergiou@unomaha.edu}).}
      Francisco R. Ortega,\thanks{Francisco R. Ortega is with the Colorado State University (email: \href{mailto:fortega@colostate.edu}{fortega@colostate.edu}).}
and Spyridon Mastorakis\thanks{Spyridon Mastorakis is with the University of Nebraska, Omaha (email: \href{mailto:smastorakis@unomaha.edu}{smastorakis@unomaha.edu}).}
}

\author{
\IEEEauthorblockN{Md Washik Al Azad\IEEEauthorrefmark{1},
Susmit Shannigrahi\IEEEauthorrefmark{2},
Nicholas Stergiou\IEEEauthorrefmark{1},
Francisco R. Ortega\IEEEauthorrefmark{3},
Spyridon Mastorakis\IEEEauthorrefmark{1}
}
\IEEEauthorrefmark{1}University of Nebraska at Omaha, \IEEEauthorrefmark{2}Tennessee Tech University, \IEEEauthorrefmark{3}Colorado State University
}

\makeatletter
\patchcmd{\@maketitle}
  {\addvspace{0.5\baselineskip}\egroup}
  {\addvspace{-10\baselineskip}\egroup}
  {}
  {}
\makeatother

\begin{document}

\maketitle


\makeatletter
\g@addto@macro{\UrlBreaks}{\UrlOrds}
\makeatother

\setlength{\skip\footins}{0.17cm}

\begin{abstract}

In today's era of Internet of Things (IoT), where massive amounts of data are produced by IoT and other devices, edge computing has emerged as a prominent paradigm for low-latency data processing. However, applications may have diverse latency requirements: certain latency-sensitive processing operations may need to be performed at the edge, while delay-tolerant operations can be performed on the cloud, without occupying the potentially limited 
edge computing resources. To achieve that, we envision {\susmit an} environment where computing resources are distributed across edge and cloud offerings. In this paper, we present the design of \sol (\textit{CL}oud + \textit{EDGE}), an information-centric hybrid cloud-edge framework, aiming to maximize the on-time completion of computational tasks offloaded by applications with diverse latency requirements. 
The design of \sol is motivated by the networking challenges that mixed reality researchers face. 
Our evaluation demonstrates that \sol can complete on-time more than 90\% of offloaded tasks with modest overheads. 


\end{abstract}






\section{Introduction}
\label{sec:intro}





Over the last few years, we have witnessed an explosion of the number of Internet of Things (IoT) devices and the amounts of data that these devices produce. The number of IoT devices is expected to grow further in the future, reaching 75 billion connected IoT devices by 2025~\cite{cisco}. This calls for pervasive edge computing deployments~\cite{shi2016edge}, where computing resources are available at the network edge for the low-latency processing of data generated by IoT and other user devices. However, considering the potentially small-scale deployments of computing resources at the network edge, it is critical that computational tasks offloaded by user devices are executed based on the latency they can tolerate by resources located at a proper distance from the users. For example, delay-sensitive 
tasks $must$ be executed as close to users as possible, while delay-tolerant tasks can be executed further away (possibly on the cloud), ensuring that: (i) the latency requirements of applications that offload the tasks are met; and (ii) resources are utilized efficiently (\eg delay-tolerant tasks do not occupy resources close to users that may be critical for the execution of delay-sensitive tasks). Furthermore, when computing resources in a particular edge network are fully utilized, user/application-offloaded tasks should be distributed in an adaptive, swift manner to nearby edge networks or a cloud depending on the latency they can tolerate.

To achieve flexible data processing and distribution of tasks, we envision hybrid computing environments where computing resources will be distributed across several edge networks and cloud offerings. In such environments, computing resources of different access latency and capacities will be available to users in a hierarchical manner. At the network edge, limited resources will be available close to users, while a vast amount of resources will be further away on the cloud at the cost of higher communication latency. As a realization of our vision, in this paper, we present \sol (\textit{CL}oud + \textit{EDGE}), an Information-Centric framework for hybrid cloud-edge computing. The \sol design is motivated by the networking challenges identified through a survey among researchers in the mixed reality community. 
\sol uses Named Data Networking (NDN) to: (i) realize a two-tier, flexible synchronization process for the exchange of resource utilization information among Edge Nodes (\ENs) within the same or different edge networks; and (ii) seamlessly distribute offloaded tasks for execution towards \ENs within the same or different edge networks or towards cloud offerings.

Our contributions are the following: (i) we motivate the \sol design through a survey conducted among mixed reality researchers in order to better understand their networking challenges and requirements; (ii) we present the design of \sol, a hybrid cloud-edge framework over NDN, to tackle the challenges of not only mixed reality applications, but also any application that offloads computational tasks with disparate latency requirements; and (iii) we perform an evaluation study of \sol and compare its performance with several baseline approaches. 
Our evaluation results demonstrate that {\fgcs \sol seamlessly integrates edge and cloud computing resources}. Specifically, \sol achieves on-time task completion rates of at least 90\% under both light and heavy load conditions with reasonable overheads, outperforming all baseline approaches by 7-78\% in terms of on-time task completion rates.

\section{Background and Prior Work}
\label{sec:relwork}


\subsection {Named Data Networking}

Named Data Networking (NDN)~\cite{zhang2014named} 
utilizes application-defined 
hierarchical naming for data publication and communication. Consumer applications send requests for ``named data", called \emph{Interest} packets, which are forwarded towards data producers based on their names. Once a producer 
receives an Interest, it sends back a \emph{Data packet} that is cryptographically signed by its producer and contains the requested content. 
To forward Interests towards producers and Data packets back to consumers, NDN forwarders 
maintain three main data structures: (i) a Forwarding Information Base (FIB), which contains entries of name prefixes along with one or more outgoing interfaces for Interest forwarding; (ii) a Pending Interest Table (PIT), which stores information about forwarded Interests that have not retrieved data yet; and (iii) a Content Store (CS), which caches retrieved Data packets to satisfy future requests for the same data.

\subsection {Cloud and Edge Computing Research}

The community has explored cloud computing approaches over NDN, often in combination with other next-generation networking technologies, such as Software Defined Networking and Network Function Virtualization~\cite{ravindran2013towards}. Named Function Networking (NFN)~\cite{sifalakis2014nfn} attempted to 
utilize lambda expressions to formulate computations and distribute them for execution to computing resources. NFaaS extended the NFN design by placing computing functions in the network and executing them through lightweight virtual machines~\cite{krol2017nfaas}. RICE augmented the capabilities offered by both NFN and NFaaS to enable consumer authentication and input parameter passing
~\cite{krol2018rice}. 
Amadeo \textit{et al.}~\cite{amadeo2016ndne}, Krol \textit{et al.}~\cite{krol2019compute}, and Mastorakis \textit{et al.}~\cite{mastorakis2019towards} have explored edge computing frameworks in NDN. Such frameworks can support the execution of programs/tasks in diverse environments, including the edge of the network.


{\fgcs Other approaches have proposed hybrid computing models based on Software Defined Networking (SDN)~\cite{8892518}, however, the SDN controller becomes a single point of failure, while approaches to replicate/distribute the controller may result in considerable overheads~\cite{mastorakis2020icedge}. Analytical modeling studies of hybrid computing systems have also been conducted~\cite{loghin2019towards}. In general, traditional IP-based solutions require complex configuration and maintenance of the communication infrastructure. On the contrary, NDN communication is name based. Nodes and networks can be added or removed transparently to the users without additional configuration or maintenance.


None of the prior works has focused on effectively distributing computational tasks over NDN to available computing resources scattered across edge networks and the cloud with the objective of satisfying the maximum possible number of tasks, each with its own completion deadline.
The initial aspiration for \sol started from understanding the requirements of a specific target community (\ie the mixed reality community). In the course of designing \sol, we realized that such requirements apply not only to mixed reality applications but, in general, to real- or near real-time applications. To this end, \sol can accommodate the requirements of diverse application use-cases, such as smart homes, public safety, industrial control systems, and {\susmit IoT applications}.}




\section{The Use-Case of Mixed Reality}
\label{sec:survey}

\begin{figure*}[!ht]
\centering
  \begin{subfigure}[t]{0.37\textwidth}
    \includegraphics[width=0.98\linewidth]{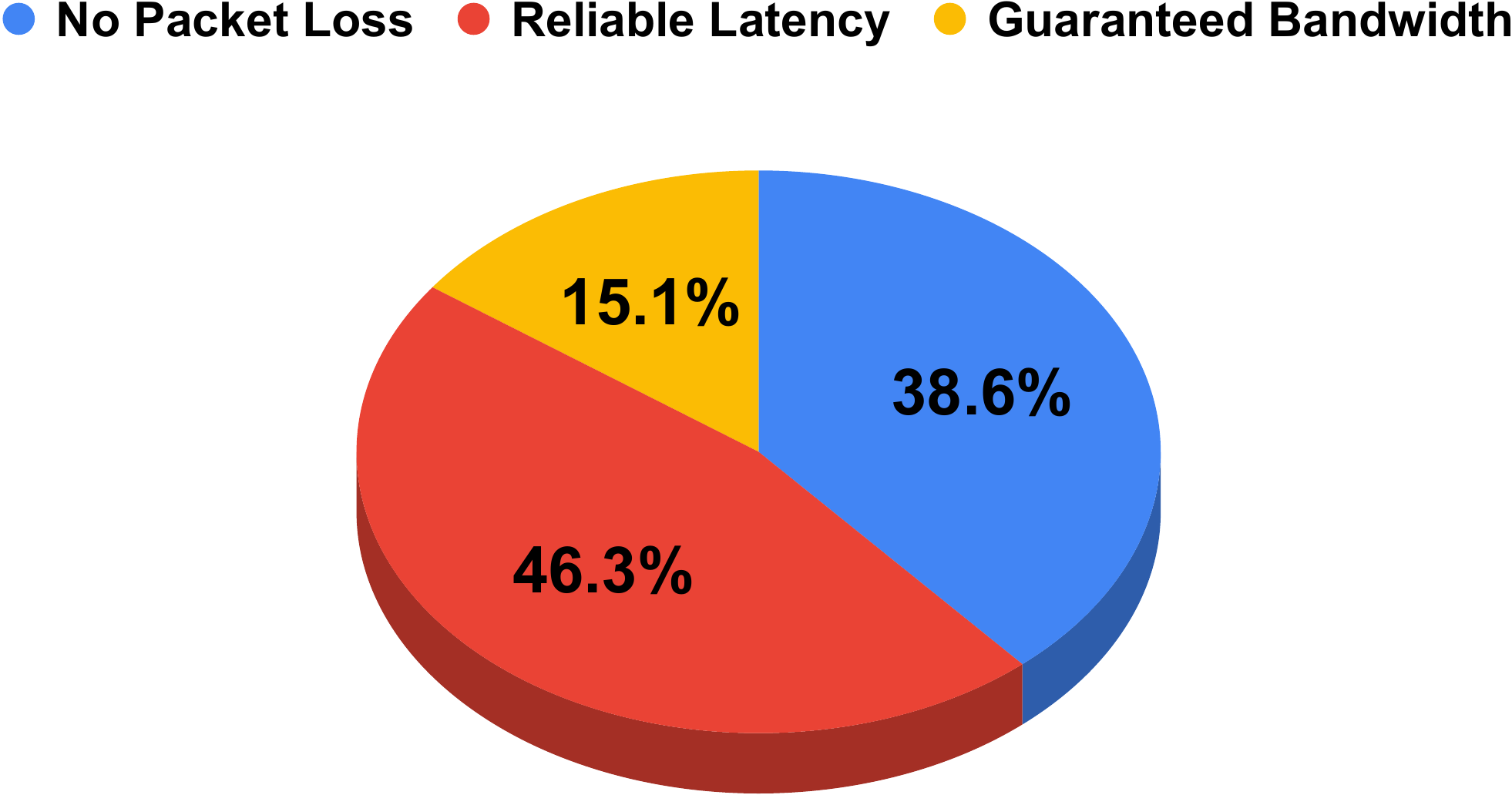}
    \caption{Networking requirements identified by mixed reality researchers}
    \label{fig:requirements}
    \end{subfigure}
  \begin{subfigure}[t]{0.25\textwidth}
  \includegraphics[width=0.98\linewidth]{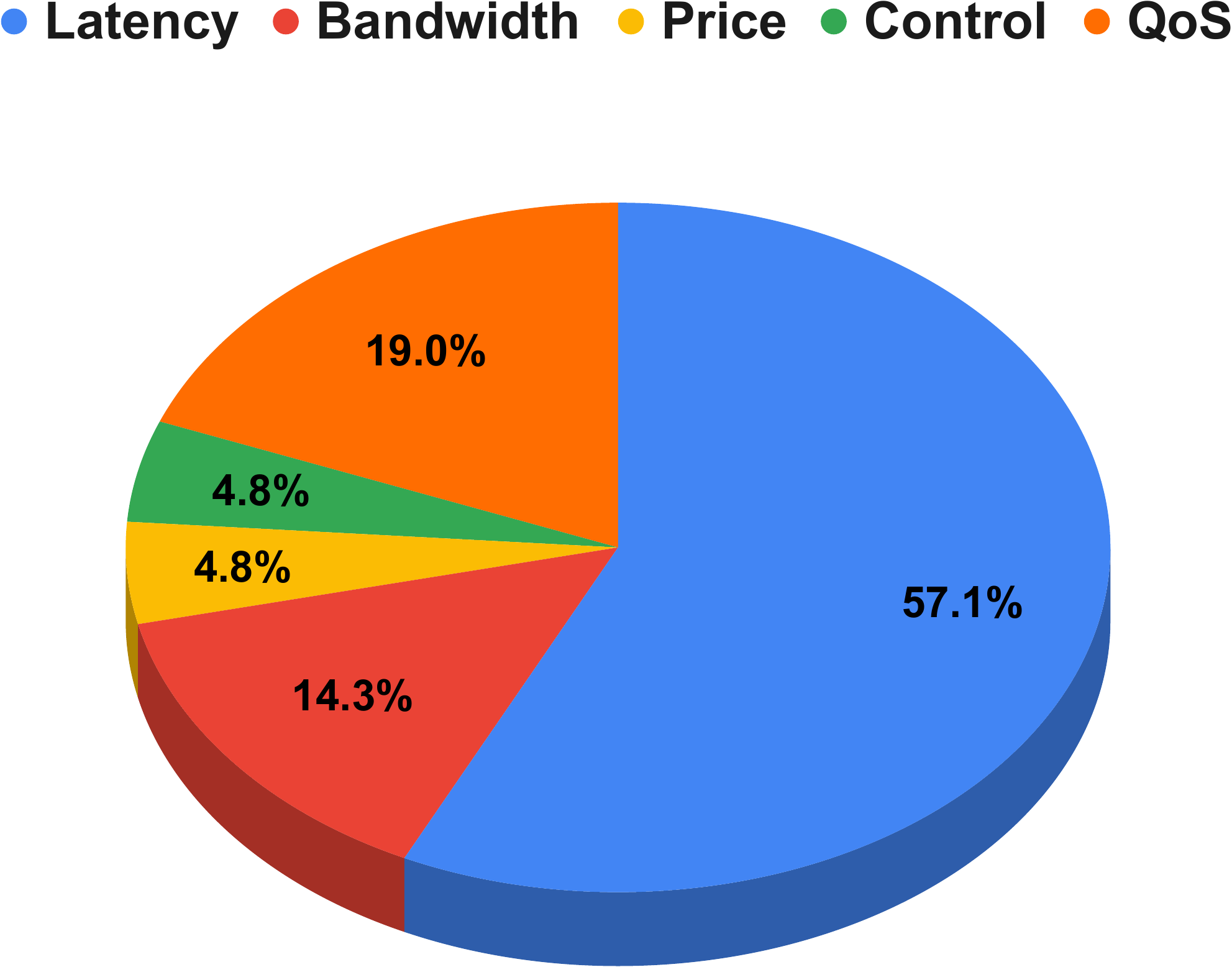}
    \caption{Networking challenges identified by mixed reality researchers}
    \label{fig:problems}  
  \end{subfigure}
  \begin{subfigure}[t]{0.30\textwidth}
  
      \includegraphics[width=0.98\linewidth]{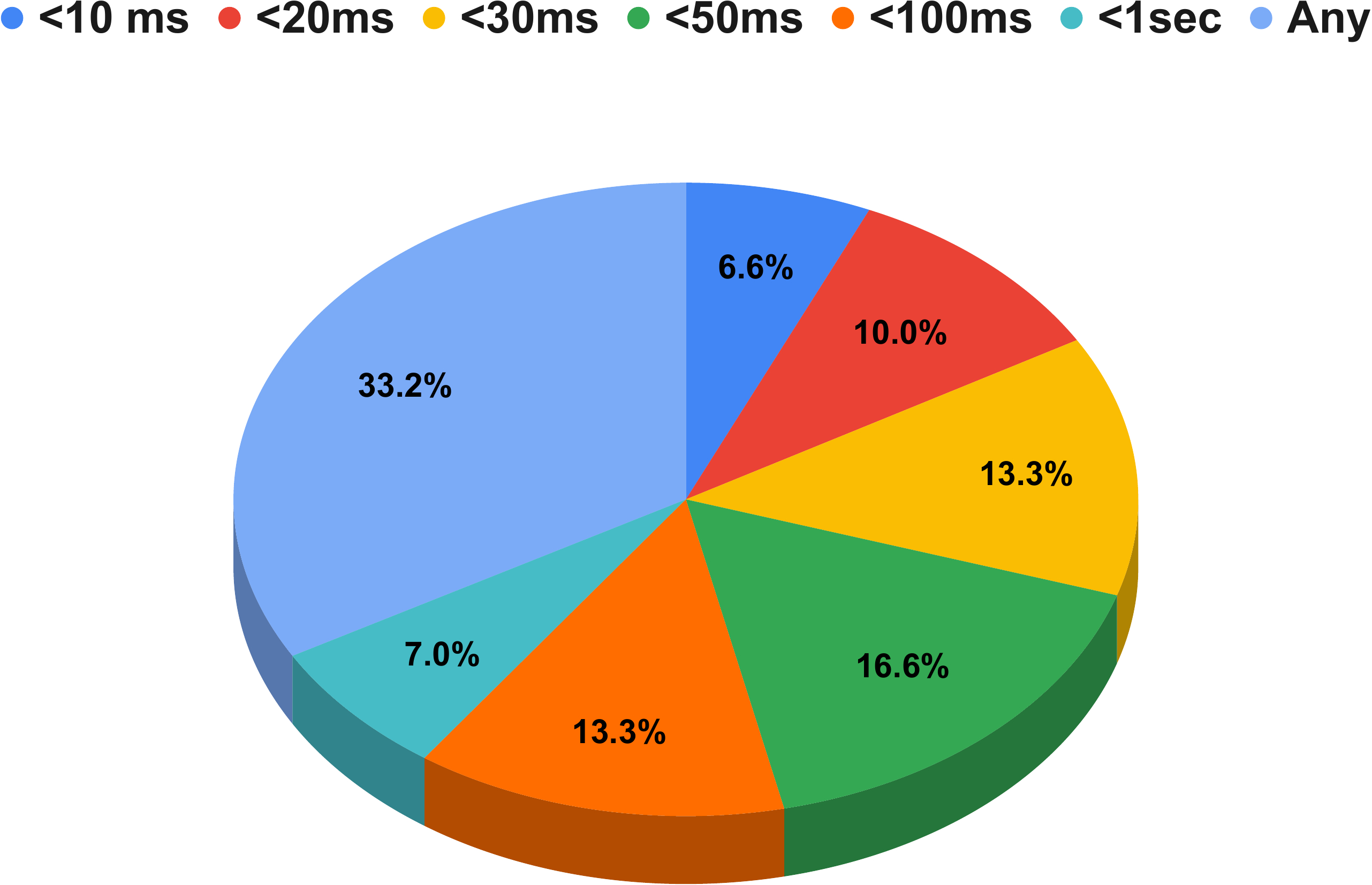}
    \caption{Tolerable end-to-end latency as identified by mixed reality researchers} 
    \label{fig:latency}
  \end{subfigure}

 \centering
 \vspace{+0.2cm}
 \caption{Mixed reality community survey results}
 \vspace{-0.1cm}
 \end {figure*}

To better understand the needs of the mixed reality community and align the design of \sol with our primary use-case, we conducted a community survey among mixed reality researchers (N=27). Figure~\ref{fig:requirements} presents the requirements of the mixed reality community in terms of networking. 46\% of the participants pointed out that latency is the most critical factor. 
A ``reliable" and consistent latency may be also needed, while having low latency at the beginning of communication followed by higher latency later on is problematic. 38\% of the participants pointed out the minimization of packet loss as their primary requirement. However, multiple respondents indicated that for the use cases where latency is important, the reduction of packet loss cannot come at the cost of increased latency. 15\% mentioned that guaranteed bandwidth is the most important requirement for their applications. 




Figure~\ref{fig:problems} 
shows the problem areas. 57\% of the responses indicated that the network struggles to meet their latency requirements. They mentioned several reasons for higher latency than what their applications can tolerate: high latency towards remote cloud offerings, inconsistent latency over production networks, and more. 
19\% of the participants suggested that they are concerned about the Quality of Service (QoS), such as low packet loss, guaranteed deadlines for the delivery of processing results, and low jitter. Only 14\% of the responses noted a lack of bandwidth as a cause for concern.



Figure~\ref{fig:latency} illustrates the latency breakdown, as desired by the community. About 46\% of the participants require latency below 50ms, while about 60\% require latency below 100ms. The participants indicated that achieving sub-100ms latency with cloud computing is unlikely, thus they often deploy and manage computing resources in their local networks. While low latency requirements have been reported previously~\cite{8329628, driscoll2017enabling}, the surprising finding in this survey was the breadth of use-cases, and how diverse the latency requirements are. Some applications indeed need ultra-low latency (less than 10-30ms). However, other applications (e.g., 2D Augmented Reality) can tolerate up to 100ms or up to a second of latency.
The diverse latency requirements of mixed reality applications cannot be accommodated through exclusive edge or cloud offerings. For example, applications that can tolerate up to 50ms of latency are unlikely to operate properly with cloud offerings, where Round-Trip Times (RTTs) may vary from 50ms to 300ms~\cite{CDNDelay}.

To address these issues, we propose \sol, a hybrid cloud-edge environment that shows promise to fulfil the diverse latency requirements of mixed reality applications. In \sol,
hierarchically distributed computing resources may be located at different distances from the users: (i) in edge networks either one hop (accessed through direct links, such as LTE/5G) or 2-3 hops away from users; and (ii) on remote clouds. 
\sol enables the execution of computational tasks with diverse latency requirements by finding the appropriate execution locations at the edge or on the cloud based on the latency that the tasks can tolerate.



\section {System Model \& Assumptions}
\label{sec:system-model}


We define an edge network as an autonomous network of Edge Nodes (\ENs), $EN_1$, $EN_2$, ..., $EN_n$, that offer a set of services (\eg object recognition, face detection) to users. The \ENs are server-class nodes with computing and storage resources. We assume that \ENs can be accessed though direct links (\eg LTE, 5G, WiFi) or links of 2-3 network hops, and that user devices (\eg mobile phones, AR headsets) are associated with an edge network within their communication range. Applications running on user devices offload computational tasks to \ENs in the edge network they are associated with by specifying the services to be invoked along with input data. The \ENs execute these tasks and return the results to the users.


We assume that multiple edge networks may be available, each administrated by the same or different entities (\eg service providers). Given that edge computing resources may be limited at any given time, we utilize an adaptive distribution scheme for offloaded tasks. When a user offloads a task, this task will be distributed for execution to available computing resources at an appropriate distance from the user based on the latency that the task can tolerate. Furthermore, when a user offloads a task but no resources are available in an edge network $x$ (\ie the \ENs of $x$ are fully utilized), $x$ can further offload (distribute) the task to a nearby edge network $y$ with available resources. When resources are not available at the edge (\ie neighboring edge networks do not have adequate resources available), resources on a cloud may be utilized.

\begin{figure}[ht]
    \centering
    \includegraphics[width=0.65\columnwidth]{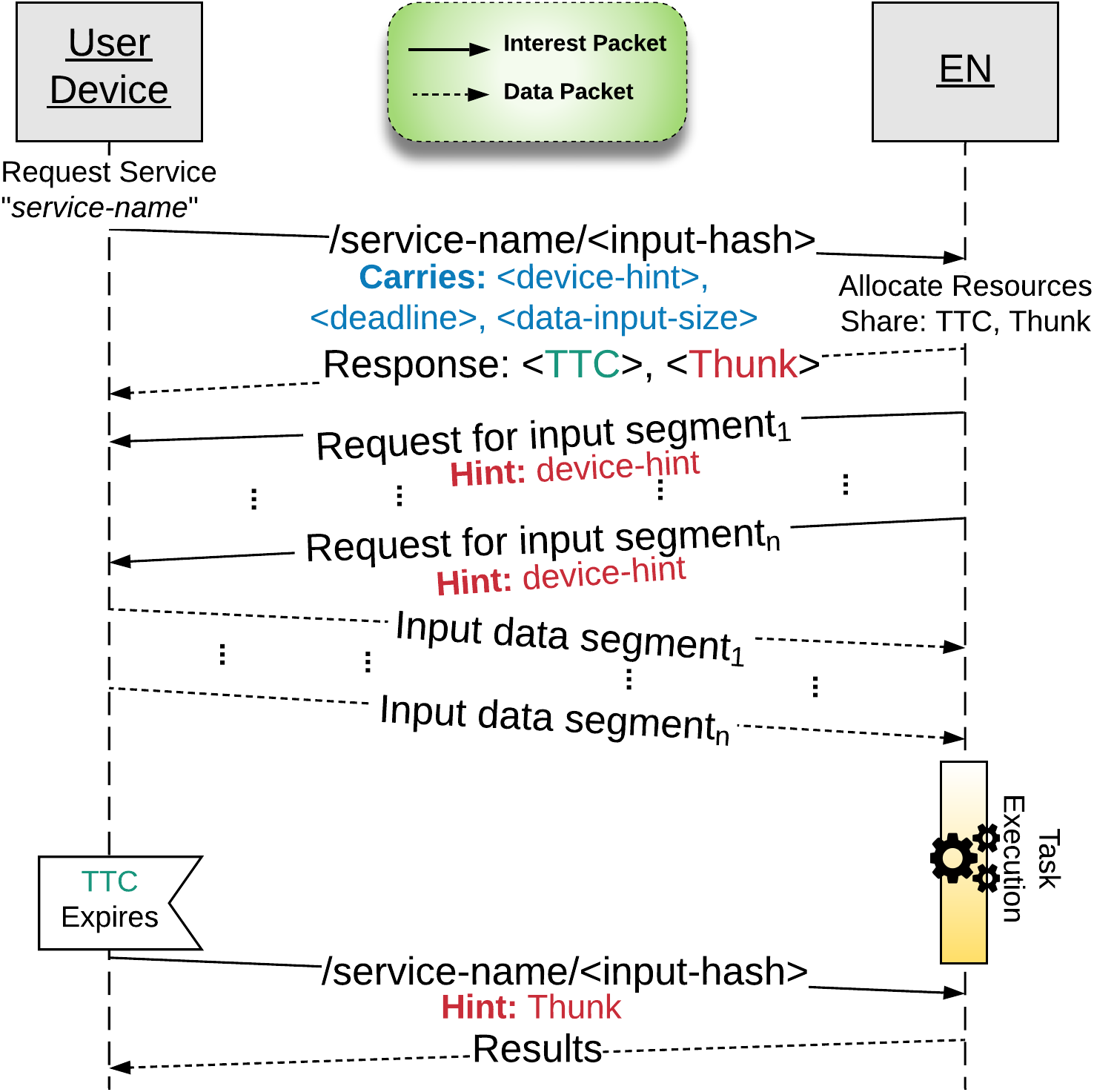}
    \vspace{-0.1cm}
    \caption{Example of the task offloading process.} 
    \label{fig:invocation}
\end{figure}

\noindent \textbf{Task naming and composition:} {\spyros Following approaches proposed in prior work~\cite{krol2018rice, mastorakis2020icedge}, we represent computational tasks as Interests with the following name format: \name{/<service-name>/<input-hash>}. An example of a task name is illustrated in Figure~\ref{fig:fig1}. The first name component of a task specifies the service to be invoked, while the second one refers to a hash of the task input data, which distinguishes tasks for the same service but with different input data. Each task is associated with a deadline by which the edge (or the cloud) needs to execute the task and return the results back to the user (\ie the delay that the user application can tolerate until it receives the task execution results). This deadline will be attached to the parameters of an Interest~\cite{ndn2019ndn} in order to leverage in-network caching for tasks with the same input. 

Input data of small sizes can be directly attached to the parameters~\cite{ndn2019ndn} of an Interest~\cite{mastorakis2020icedge}. As illustrated in Figure~\ref{fig:invocation}, additional Interest-Data packet exchanges may be employed to pass input data of larger sizes (\eg high-resolution images or video frames)~\cite{krol2018rice}. In such cases, the \EN will utilize the forwarding hint of the user device~\cite{ndn2015ndn}\footnote{An Interest can carry both a forwarding hint and a name. The forwarding hint is a name identifier that specifies ``where" (\eg to which \EN or user device) an Interest for a certain service or data (identified in the Interest name) should be forwarded.} 
to request the input data from the device. The \EN will also send to the device an estimated Time To Completion (TTC) for the offloaded task and a thunk~\cite{ingerman1961way}, which is a name that will allow the device to reach the particular \EN that executes the task after TTC has expired and retrieve the task execution results. A thunk may consist of a concatenation of the name prefix of the \EN executing the task and a hash that represents the internal state of execution and identifies the execution of a specific task.

}

\section{Design}
\label{sec:design}

\vspace{-0.2cm}



\subsection {Design Overview}

\begin{figure}[t]
    \centering
    \includegraphics[width=0.8\columnwidth]{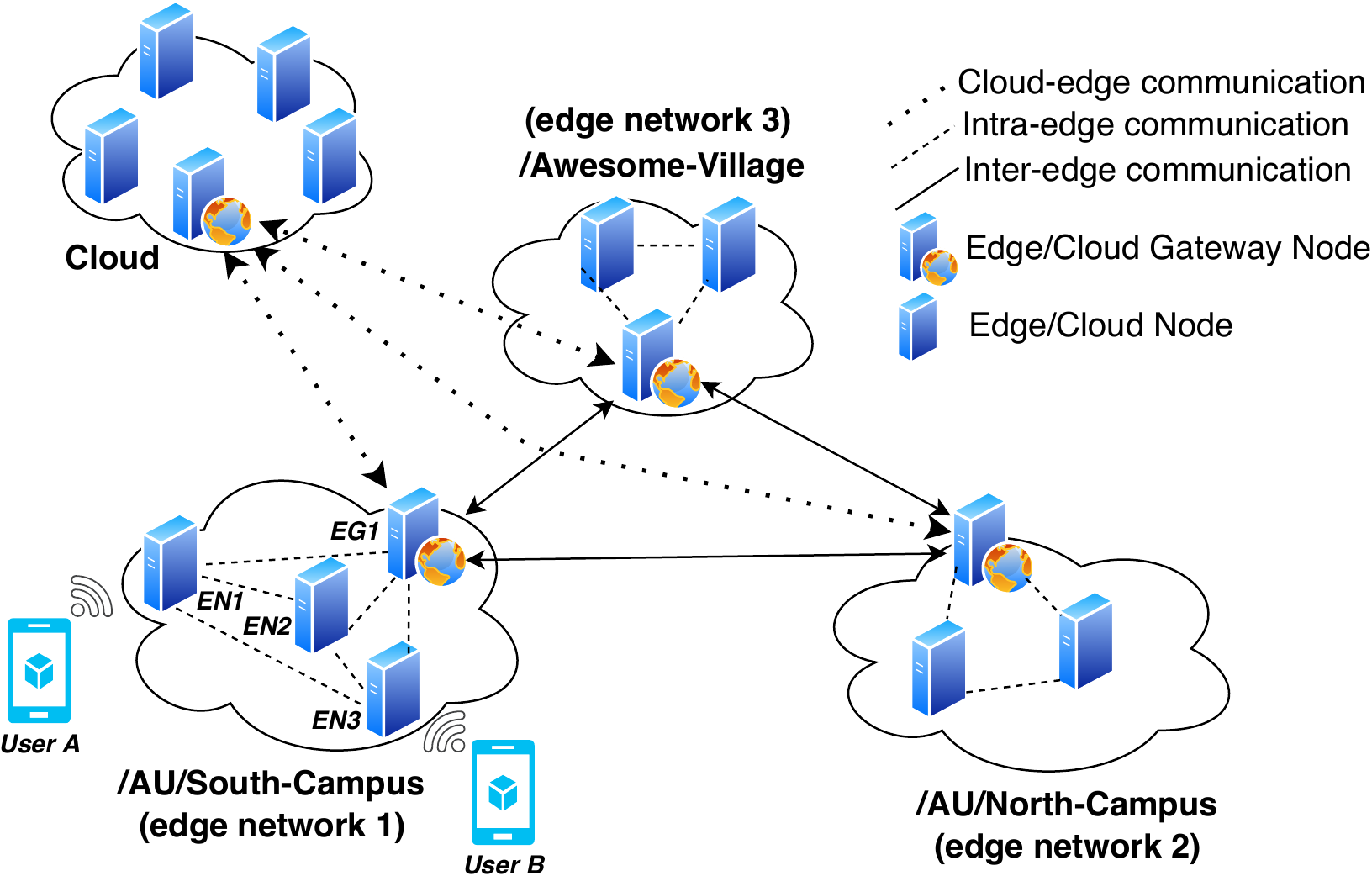}
    \vspace{-0.15cm}
    \caption{\sol running example. Three edge networks are available with a connection to a remote cloud. One edge network is located on the south campus of the Awesome University (AU), one on the north campus of AU, and one in the awesome village where AU is located.}
    \label{fig:system_architecture}
\end{figure}

We present the design of \sol through a running example (Figure~\ref{fig:system_architecture}), where multiple edge networks and a remote cloud are available to execute tasks offloaded by users. Edge networks are interconnected through a network of Edge Gateways (\ONs). Specifically, one of the \ENs in each edge network is designated as an \ON. 
\sol features a two-tier synchronization process: the first tier takes place within an edge network and involves the \ENs and the \ON in this network, while the second tier involves synchronization among the \ONs of different edge networks. Through this process, we enable: (i) \ENs to be aware of the up-to-date availability of computing resources at all other \ENs in their edge network; and (ii) \ONs to be aware of the up-to-date availability of computing resources across edge networks. In addition to resource availability, \ENs will be able to estimate the RTT to other \ENs in the same edge network, while \ONs will be able to estimate the RTT to other edge networks. 

{\susmit Our synchronization process is not bound to a specific NDN synchronization protocol, but is able to employ} existing NDN synchronization protocols~\cite{li2019distributed}, involving two tiers of synchronization to scale up the overall process and mitigate the resulting overhead as the number of edge networks and \ENs increases. At the same time, the different synchronization tiers enable a flexible design, where the synchronization frequency of the first tier is decoupled from the synchronization frequency of the second tier. Moreover, the synchronization frequency of \ENs in an edge network can be modified independently of the frequency of other edge networks, effectively adapting to the ongoing operational conditions. For instance, more frequent synchronization can be selected when user traffic and resource utilization change rapidly, while less frequent synchronization can be selected when conditions are stable. We evaluate the impact of the synchronization frequency on the performance and overhead of \sol in Section~\ref{sec:eval}.






\sol adaptively distributes offloaded tasks for execution with the objective of maximizing the task satisfaction rate (\ie the execution results will be returned to users by the deadline of each task). Overall, tasks will be executed as further from (or closer to) the users as their deadlines allow. To meet this goal, \sol ensures that delay-sensitive tasks will find available resources at \ENs close to users, while tasks that can tolerate additional delay will be executed at \ENs further away from users or even on the cloud. For example, once a user offloads a delay-sensitive task to an \EN in edge network 1 (Figure~\ref{fig:system_architecture}), the \EN tries to execute the task if it has available resources. If this \EN has no available resources, it offloads the task to the closest (in terms of RTT) \EN with enough resources to satisfy the deadline. 
If no \ENs or the \ON in edge network 1 have available resources and the task cannot tolerate execution on the cloud, the task is forwarded through the network of \ONs to another edge network with available resources (\eg edge network 2 or 3 in Figure~\ref{fig:system_architecture}). In the case of a delay-tolerant task, the task is distributed to the furthest \EN within an edge network (or even the cloud) that can satisfy the task deadline.

To achieve adaptive and accurate distribution of tasks based on the latency they can tolerate, \ENs need to be aware of the network delay to available computing resources as well as the time that these resources may need to execute the tasks. To accomplish that, \sol establishes profiles for each service offered in an edge network. These profiles help \ENs to estimate how much time each service needs to execute a task. \ENs in an edge network exchange such profile information (statistics) through the synchronization process within their network, while, optionally, these statistics can also be exchanged among edge networks through the \ONs. 

\subsection {Namespace Design}

\begin{figure}[t]
\centering
\begin{subfigure}[]{\columnwidth}
    \centering
   \includegraphics[width=0.5\linewidth]{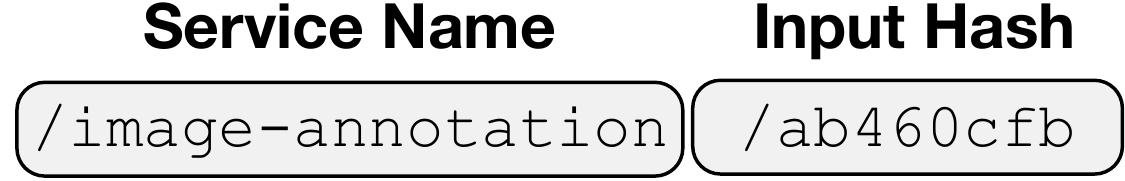}
   \caption{Task namespace}
   \label{fig:fig1} 
\end{subfigure}
\vspace{0.2cm}
\begin{subfigure}[b]{\columnwidth}
\centering
   \vspace{0.5cm}
   \includegraphics[width=0.6\linewidth]{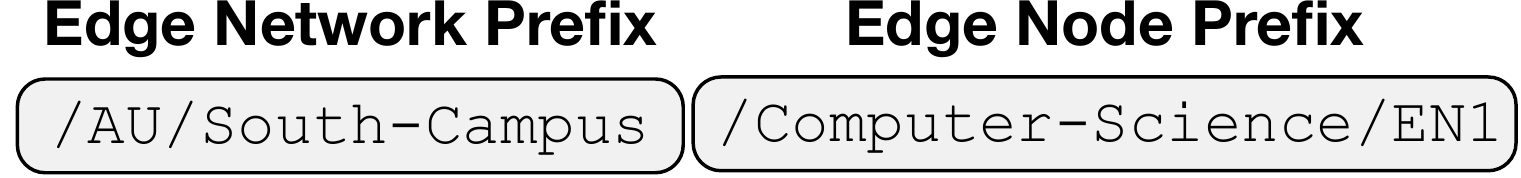}
   \caption{\EN namespace}
   \label{fig:fig2}
   \vspace{0.2cm}
\end{subfigure}
\begin{subfigure}[b]{\columnwidth}
   \centering
   \includegraphics[width=0.5\linewidth]{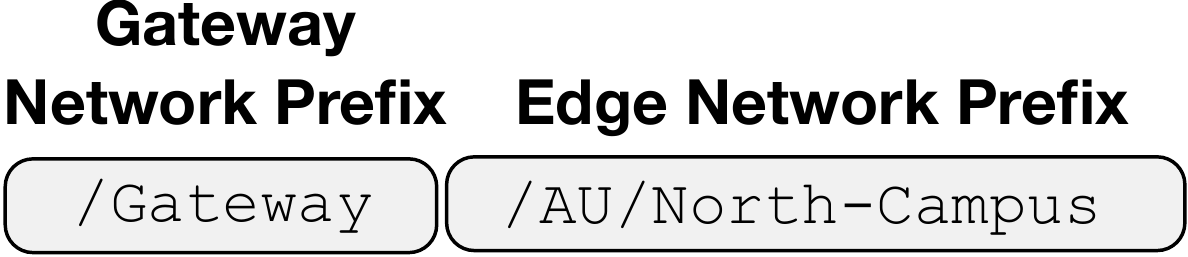}
   \caption{\ON namespace}
   \label{fig:fig3}
   \vspace{0.3cm}
\end{subfigure}

\caption{Namespace design}
\label{fig:bigfig}
\end{figure}

Figure~\ref{fig:bigfig} illustrates the namespace design in \sol. As we mentioned in Section~\ref{sec:system-model}, we represent tasks as Interests that adhere to the following naming format: \name{/<service-name>/<input-hash>}. The first name component of a task specifies the service to be invoked and the second one refers to a hash of the task input data. These Interests also carry the task completion deadline in their parameters~\cite{ndn2019ndn}, so that tasks for the same input can take advantage of NDN in-network caching. In Figure~\ref{fig:fig1}, we illustrate the name of a task that invokes an annotation service for a certain image as an input. 

Each \EN has a name under the name prefix of the edge network it belongs to. For example, in Figure~\ref{fig:fig2}, we present the name of an \EN on the south campus of the Awesome University (AU) 
and, specifically, in the Computer Science department. Furthermore, each \ON will have a name prefix for communication with other edge networks. 
For example, for the execution of tasks offloaded on the south campus of AU by computing resources on the north part of the campus, the tasks will be distributed from the \ON of the edge network on the south campus towards the \ON of the edge network on the north campus through the name presented in Figure~\ref{fig:fig3}.

\subsection {Distribution of Tasks Within an Edge Network}


\noindent \textbf{Synchronization process:} The first tier of synchronization in \sol involves the \ENs and the \ON within an edge network. This process takes place, so that the \ENs and the \ON can maintain up-to-date information needed for the adaptive and accurate distribution of offloaded tasks. Such information may include the latest utilization of the computing resources of each \EN and the latest statistics about the execution of each service that will be used for the creation and maintenance of the service profiles. During this process, each \EN can also measure the RTT to every other \EN.
{\susmit Note that an alternative to synchronization may be an event driven model where \sol finds an appropriate location of execution as tasks arrive. However, the additional lookup and task placement time may cause problems for latency sensitive tasks.}


\noindent \textbf{Building service profiles:} {\spyros Each computational task invokes a certain service. Over time, \ENs execute offloaded tasks and build execution profiles for each invoked service by collecting statistics during the execution of each task. The statistics to be collected can be determined by the network administrators, but metrics of interest {\spyros may include execution times of tasks (\eg average values, mean values, 95th percentile, or distributions)} per service. 
To take into account the heterogeneous hardware specifications that \ENs may have, service profiles can further include the execution times and the utilization of resources (\eg required CPU cores, RAM) for different hardware setups}. During the synchronization process, \ENs share with each other updates on the service profiles they have built. Through such profiles, \ENs can estimate the computing resources needed for the execution of tasks for a service and how long the execution of tasks for a service might take.


\noindent \textbf{Task distribution process:} Once a task is offloaded from a user application to an \EN, the \EN estimates whether the task should be executed right away or it should be distributed to an \EN further away from the user (lazy task execution). To estimate how far or close to the user a task should be executed, \ENs consider the following factors: (i) the task deadline; (ii) the RTT towards other \ENs in this edge network; (iii) an estimate of how long the task execution might take based on the profile of the invoked service; and (iv) the availability of computing resources on \ENs across the edge network. 

For example, in Figure~\ref{fig:system_architecture}, let us assume that user A offloads a task in edge network 1 that reaches \ENnospace1 for execution. \ENnospace1 will initially use factors (i), (ii), and (iii) to determine whether the task needs to be executed right away or it can be executed by an \EN in this network further away from the user (\ie \ENnospace2 and \ONnospace1). To avoid resource exhaustion, we avoid distributing offloaded tasks between \ENs that are reachable by users through a direct link (\eg WiFi, LTE/5G), such as \ENnospace1 and \ENnospace3 in Figure~\ref{fig:system_architecture}, unless there are no other computing resources available in an edge network.
Assuming that the task can tolerate to be executed by both \ENnospace2 and \ONnospace1, but \ONnospace1 does not have available resources, \ENnospace1 will attach \ENnospace2's prefix (\eg \name{/AU/South-Campus/Computer-Science/EN2} following the namespace design of Figure~\ref{fig:bigfig}) as the task's forwarding hint, so that the task is forwarded towards \ENnospace2. If both \ENnospace2 and \ONnospace1 have adequate resources, \ONnospace1 will be preferred, since it makes computing resources closer to users available for tasks that may not tolerate the network delay towards \ONnospace1. In either case, \ENnospace1's resources will stay available for tasks that cannot tolerate to be distributed to other \ENs for execution.

\subsection {Task Distribution Across Edge Networks and Cloud}


\noindent \textbf{Gateway synchronization process:} The second tier of synchronization happens among the \ONs. Each \ON is responsible for synchronizing with other \ONs on behalf of the \ENs in its own edge network. This process takes place, so that each \ON is aware of which edge networks have available resources and estimate how far these resources might be, so that tasks can be distributed from one edge network to another for execution when computing resources are occupied within a network. 

During the gateway synchronization process, the information to be synchronized among the \ONs can be determined by the edge network administrators. Information of interest may include: (i) the RTT towards the closest \EN in each \ON's network that has available resources (0 if the \ON itself has available resources); and (ii) the utilization of the resources of this \EN (or the \ON itself). \ONs may also optionally exchange aggregated statistics about services invoked within their networks to increase the accuracy of the established service profiles. Through the message exchanges for synchronization, the \ONs can measure the RTT to each other. 

\noindent \textbf{Task distribution across edge networks:} In contrast to the cloud, which may offer an abundance of computing resources, edge networks typically offer limited resources. When an edge network does not have available resources to execute newly offloaded tasks, such tasks will be forwarded to the \ON of the network. The \ON will determine based on the task deadline, the availability of resources in other edge networks, and the RTT towards these resources, whether the task can tolerate the delay for execution by another edge network or the cloud. Subsequently, a task will be forwarded to the cloud (preferable if the task can tolerate the delay given the abundance of cloud resources) or another edge network with the goal of meeting the task deadline. {\fgcs As a result, tasks can be executed among edge networks based on their resource availability. }

For distribution of tasks across edge networks, we utilize forwarding hints. Specifically, the sending \ON attaches the name prefix of the destination edge network (or cloud) as the forwarding hint of the original task. The name prefixes of edge networks for communication through the gateways follow the namespace design of Figure~\ref{fig:fig3}. For example, in Figure~\ref{fig:system_architecture}, let us assume that through the synchronization process within edge network 1, \ENs are aware that no resources are available in this edge network. In this case, a newly offloaded task will be forwarded by \ENs to \ONnospace1. \ONnospace1 will decide whether this task can tolerate to be executed on the cloud or it needs to be distributed to another edge network for execution. Assuming that \ONnospace1 decides to distribute the task to edge network 2, \ONnospace1 will attach the prefix \name{/Gateway/AU/North-Campus} as the forwarding hint of the original task.


\vspace{-0.15cm}

\section{Evaluation}
\label{sec:eval}

\vspace{-0.05cm}

In this section, we evaluate \sol through a simulation study. Our goal is to evaluate: (i) whether \sol can successfully meet the completion deadlines of tasks with diverse latency requirements; (ii) the overhead associated with \sol; and (iii) whether \sol can offer reliable latency to applications for the completion of their tasks. 

\subsection {Evaluation Setup}

\begin{figure}[t]
    \centering
    \includegraphics[width=0.64\columnwidth]{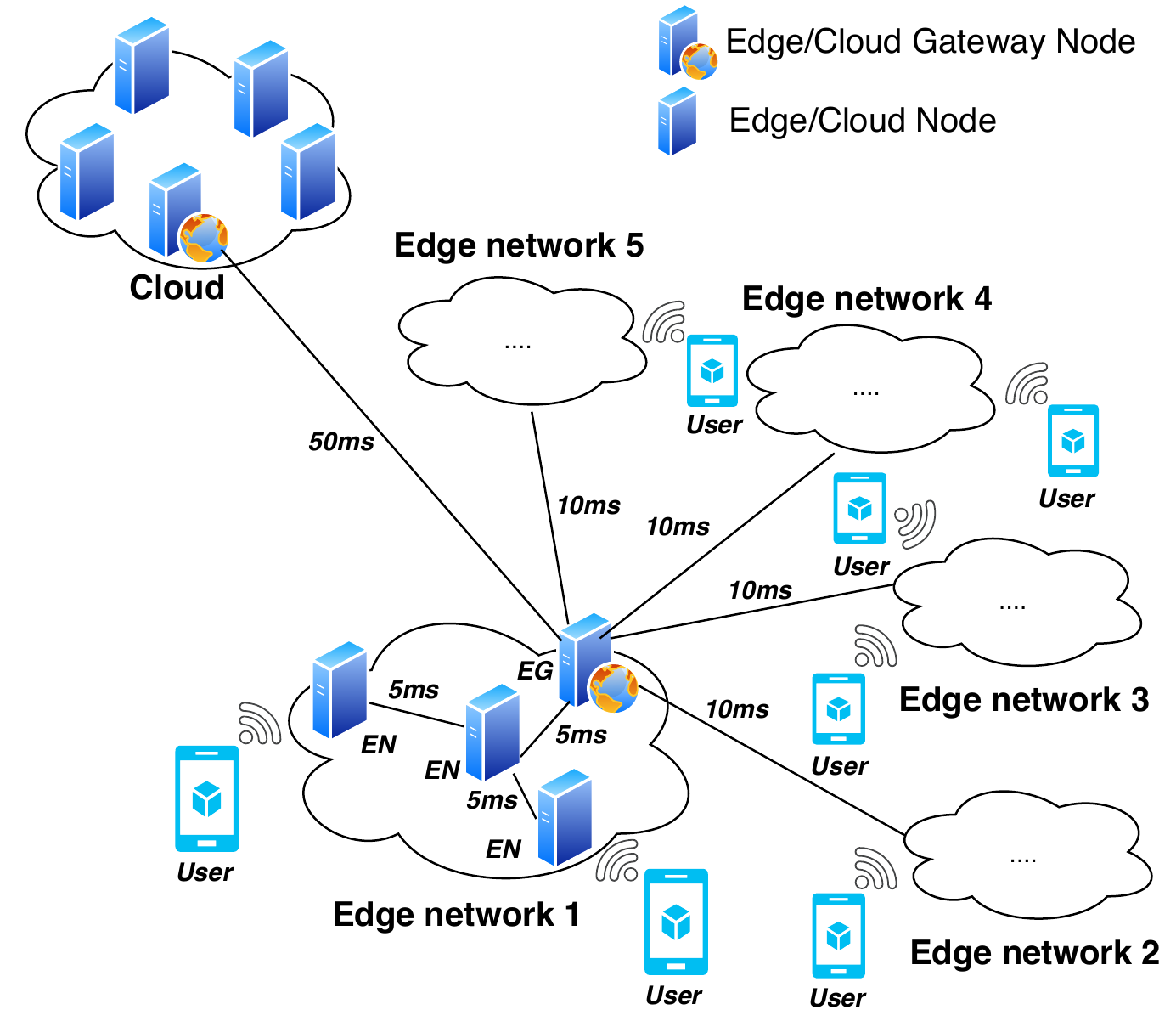}
    \vspace{-0.1cm}
    \caption{Evaluation topology. {\fgcs Note that we experimented with varying link delays and numbers of edge networks, services, and users, concluding that the results of these experiments follow the same trend as the results we present in Section~\ref{subsec:results}.}}
    \label{fig:topology}
\end{figure}

We have implemented \sol in ndnSIM~\cite{mastorakis2017evolution}. 
Figure~\ref{fig:topology} shows the topology and Table~\ref{tab:parameters} shows the simulation parameters.
Each edge network mirrors the topology and setup of edge network 1 in Figure~\ref{fig:topology}. 
To determine the one-way network latency from users to the cloud, we ran 1000 pings from various locations in the US to Amazon Web Services (AWS) servers in regions around the world using a web tool (https://www.cloudping.info)
Our measurements showed that the latency to AWS servers in different US regions varies from 40ms to 80ms, therefore, we selected the latency between each \ON and the cloud to be 50ms (about 60ms from users to the cloud). 
Each \ON is 5 hops away from the cloud, while each user is 8 hops away from the cloud. The latency and hop count values follow values reported in recent studies~\cite{nguyen2019cloud} and cloud computing trends~\cite{amazon}. 


Each user in our topology randomly selects one of the offered services. Following the conclusions of our survey (Section~\ref{sec:survey}), the services are selected from one of the following categories: (i) delay-sensitive services invoked by tasks with deadlines between 10ms and 50ms; (ii) \name{regular type} services invoked by tasks with deadlines between 50ms and 100ms; and (iii) delay-tolerant services invoked by tasks with deadlines between 100ms and 1000ms. Each service is associated with a deadline selected based on its category. For example, a delay-sensitive service $s_1$ will be associated with a deadline randomly selected between 10ms and 50ms and the tasks that invoke $s_1$ will have the associated deadline. We experimented with two load profiles: (i) \emph{light load:} each user offloads 2 to 8 tasks per second for a total of about 500 tasks per second; and (ii) \emph{heavy load:} each user offloads 10 to 30 tasks per second for a total of about 2,000 tasks per second. We also implemented a mechanism for two-tier synchronization, where each tier synchronizes periodically and independently of the other. In Section~\ref{subsec:results}, we present the average results collected over 10 runs for a total of 500,000 tasks per run.

We compare \sol to the following baseline approaches: (i) \emph{cloud-only:} tasks are exclusively offloaded onto the cloud for execution; (ii) \emph{edge-only:} tasks are offloaded to \ENs for execution. If an \EN does not have available resources, it buffers incoming tasks for later execution in a first come first served manner once resources become available; (iii) \emph{cloud-edge:} tasks are initially offloaded to \ENs for execution. If an \EN does not have available resources, then the tasks are sent to the cloud for execution; and (iv) \emph{adaptive cloud-edge:} tasks are offloaded to \ENs for execution. If an \EN does not have available resources, tasks are distributed to the closest (in terms of RTT) \EN. The main difference with \sol is that adaptive cloud-edge does not consider how much latency a task can tolerate when decisions about the task distribution are made. If no available resources exist within an edge network, tasks will be forwarded through an \ON to the closest (in terms of RTT) edge network with available resources. If none of the edge networks has available resources, tasks will be sent to the cloud. Our evaluation metrics include the following:

\begin{table}[t]
\vspace{+0.25cm}
\caption{Simulation parameters.} 
\label{tab:parameters}
\resizebox{\columnwidth}{!}{%
\tiny
\begin{tabular}{|c|c|}
\hline
\textbf{Parameter}                                                                                     & \textbf{Value(s)}                                               \\ \hline
Number of edge networks                                                                                & 5                                                               \\ \hline
Number of users                                                                                        & 100                                                             \\ \hline
Number of services                                                                                     & 50                                                              \\ \hline
Network Stack & \begin{tabular}[c]{@{}c@{}}NDN directly on top of the MAC layer (IEEE 802.11n \\ for wireless and IEEE 802.3 for wired connections)\end{tabular} \\ \hline
\begin{tabular}[c]{@{}c@{}}Number of tasks that \ENs and \\ \ONs can execute simultaneously\end{tabular} & 8                                                               \\ \hline
\begin{tabular}[c]{@{}c@{}}Total number of offloaded\\ tasks per simulation run\end{tabular}           & 500,000                                                         \\ \hline
Total simulation runs                                                                                  & 10                                                              \\ \hline
Task execution times                                                                  & 40\%-60\% of task deadline (randomly selected)                                                                  \\ \hline
\end{tabular}%
}
\end{table}

\begin{itemize} 

\item \textbf{Task satisfaction rate:} The percent of tasks that are completed on-time (\ie the tasks are executed and their results are returned to users by their associated deadlines). 

\item \textbf{Normalized overhead:} The volume of traffic generated for the completion of offloaded tasks normalized by the total size of tasks. For \sol, the overhead includes the generated traffic for the two-tier synchronization mechanism and the distribution of tasks for execution within an edge network, across edge networks, or towards the cloud. For 
all other approaches, the overhead includes the traffic for the distribution of tasks for execution by the edge or cloud. 

\item \textbf{Task completion latency:} The average latency for the completion of offloaded tasks.

\item \textbf{Reliability of task completion latency:} The standard deviation of the completion times of tasks for a certain service from the deadline associated with this service.

\end{itemize}

\vspace{-0.3cm}

\subsection {Evaluation Results}
\label{subsec:results}

\begin{figure}[th]
	\centering
	\vspace{-0.4cm}
	\captionsetup[subfigure]{aboveskip=-1pt,belowskip=-1pt}
	\begin{subfigure}[b]{1\columnwidth}
		\centering
		\includegraphics[width=0.65\columnwidth]{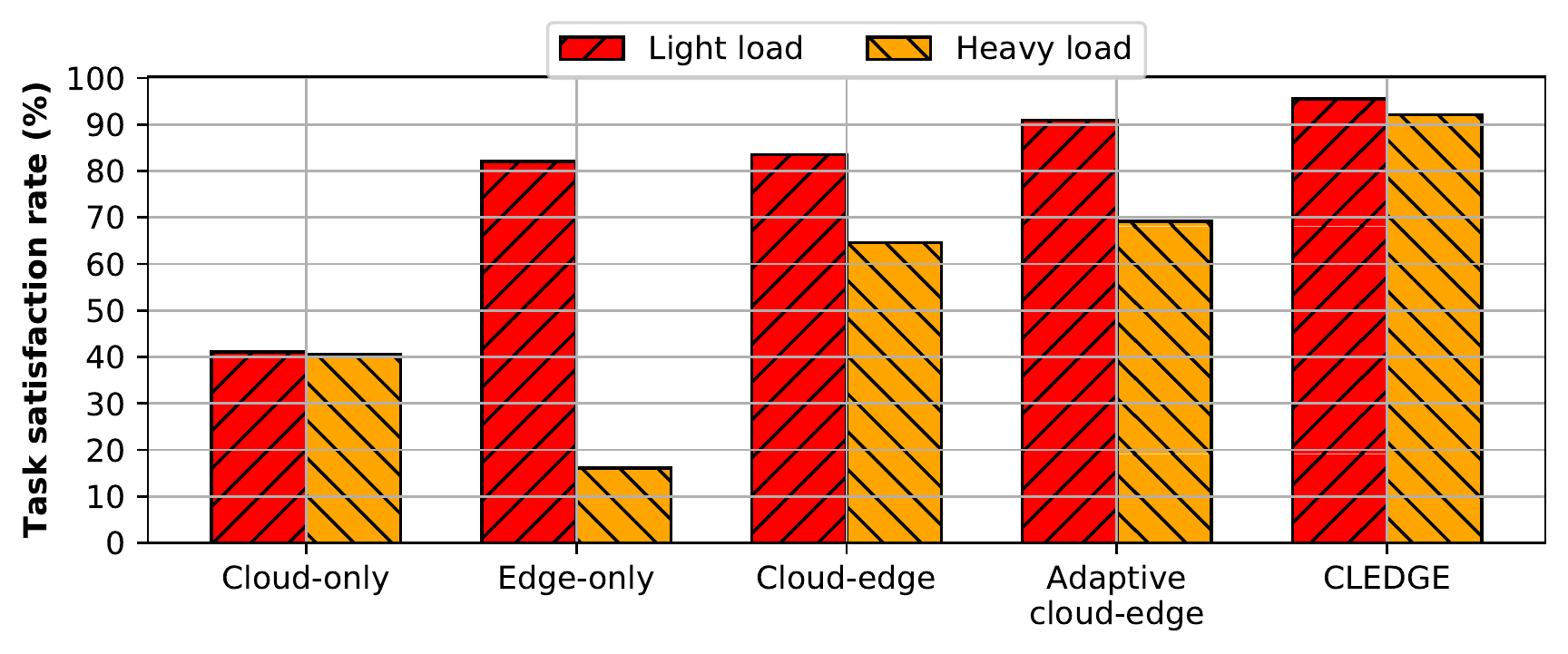}
		\subcaption{Task satisfaction rate}
		\label{Figure:satisfaction-rate}
	\end{subfigure}
	\begin{subfigure}[b]{1\columnwidth}
		\centering
		\includegraphics[width=0.65\columnwidth]{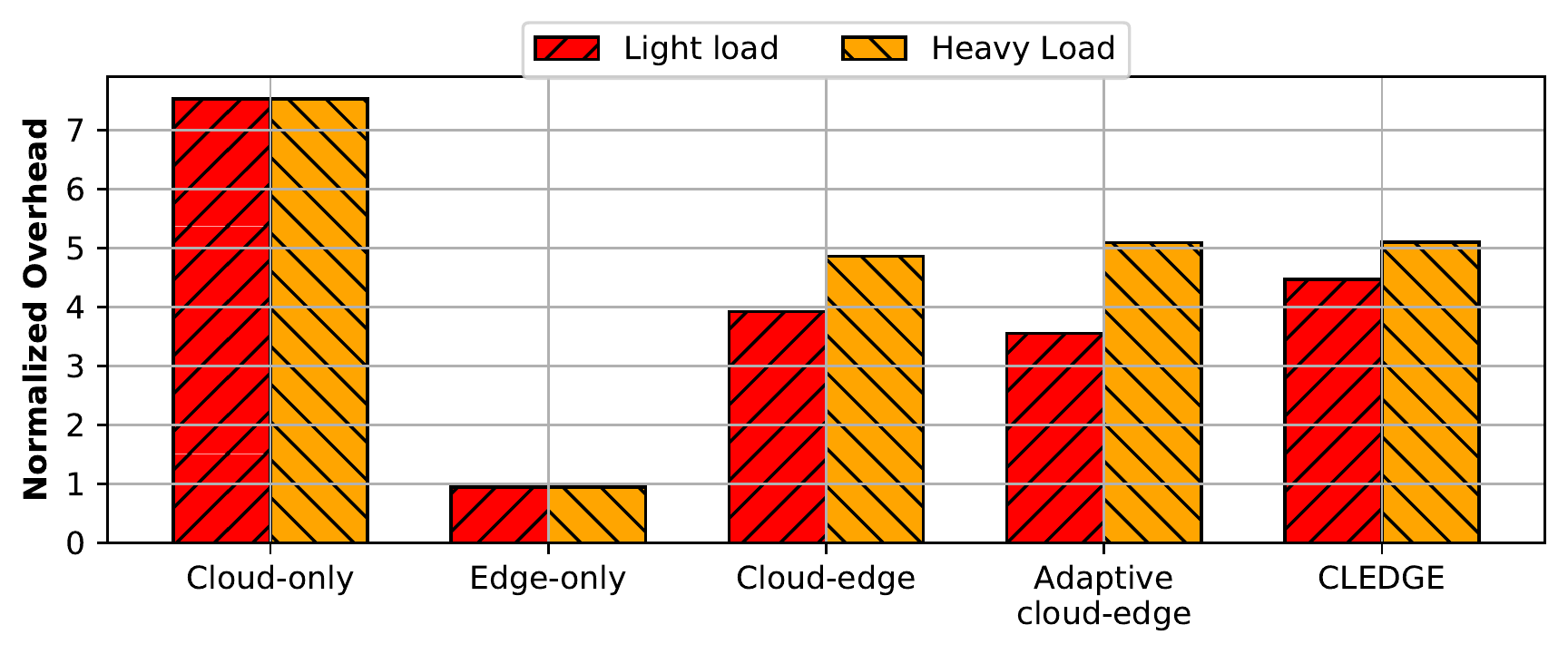}
		\subcaption{Normalized overhead}
		\label{Figure:overhead}
	\end{subfigure}
	\vspace{-0.5cm}
	\caption{Task satisfaction rate and overhead results.}
	\label{figure:comparison}
\end{figure}

\noindent \textbf{Task satisfaction rate and overhead:} In Figure~\ref{figure:comparison}, we present results for the task satisfaction rate and normalized overhead of \sol compared to other approaches. Our results indicate that the execution of tasks on the cloud (cloud-only approach) results in low task satisfaction rates and significant overhead, since tasks are forwarded far away from users. 
Execution of tasks only by the \ENs that receive them from users (edge-only approach) results in low overhead under light and heavy loads, low satisfaction rates under heavy loads (the resources of \ENs are always fully utilized), and relatively high satisfaction rates for low loads (the resources of \ENs are in general available). 
These results signify the need for a hybrid cloud-edge computing model, since exclusive execution of tasks by the cloud or edge cannot lead to satisfactory on-time completion rates under both low and high loads. Furthermore, cloud-edge and adaptive cloud-edge result in reasonable overhead (between the range of the cloud-only and edge-only overhead), while they achieve relatively high satisfaction rates for light loads and reasonable satisfaction rates for heavy loads. 

\sol is able to satisfy 95\% and 92\% of the offloaded tasks for light and heavy loads respectively, achieving 7-78\% higher satisfaction rates than the compared approaches. Specifically, \sol satisfies 13\% and 7\% more tasks than cloud-edge and adaptive cloud-edge respectively for light loads, while, for heavy loads, \sol satisfies 28\% and 24\% more tasks than cloud-edge and adaptive cloud-edge respectively. For light loads, the \ENs that directly receive offloaded tasks from users in general have available computing resources to execute these tasks. However, under heavy loads, the computing resources of \ENs are in general fully utilized, therefore, \sol can successfully distribute tasks based on their deadlines to available resources within the same or different edge networks or onto the cloud. 

In terms of overhead, \sol achieves reasonable overheads in the range between cloud-only and edge-only. Specifically, \sol's overhead is marginally higher (about 2-4\%) than cloud-edge and adaptive cloud-edge for high loads, while, for light loads, \sol results in about 11\% and 17\% higher overheads than cloud-edge and adaptive cloud-edge respectively. This is attributed to the fact that \sol aims to provide reliable latency to applications with diverse latency requirements under network and resource loads that may rapidly change. To this end, it ensures that a part of the computing resources of \ENs close to users will be available to execute latency-sensitive tasks that may be received in the future. This is achieved at the price of forwarding tasks that can tolerate latency to resources further away from users. 


\begin{figure}[th]
	\centering
	\captionsetup[subfigure]{aboveskip=-1pt,belowskip=-1.2pt}
	\begin{subfigure}[b]{1\columnwidth}
		\centering
		\includegraphics[width=0.73\columnwidth]{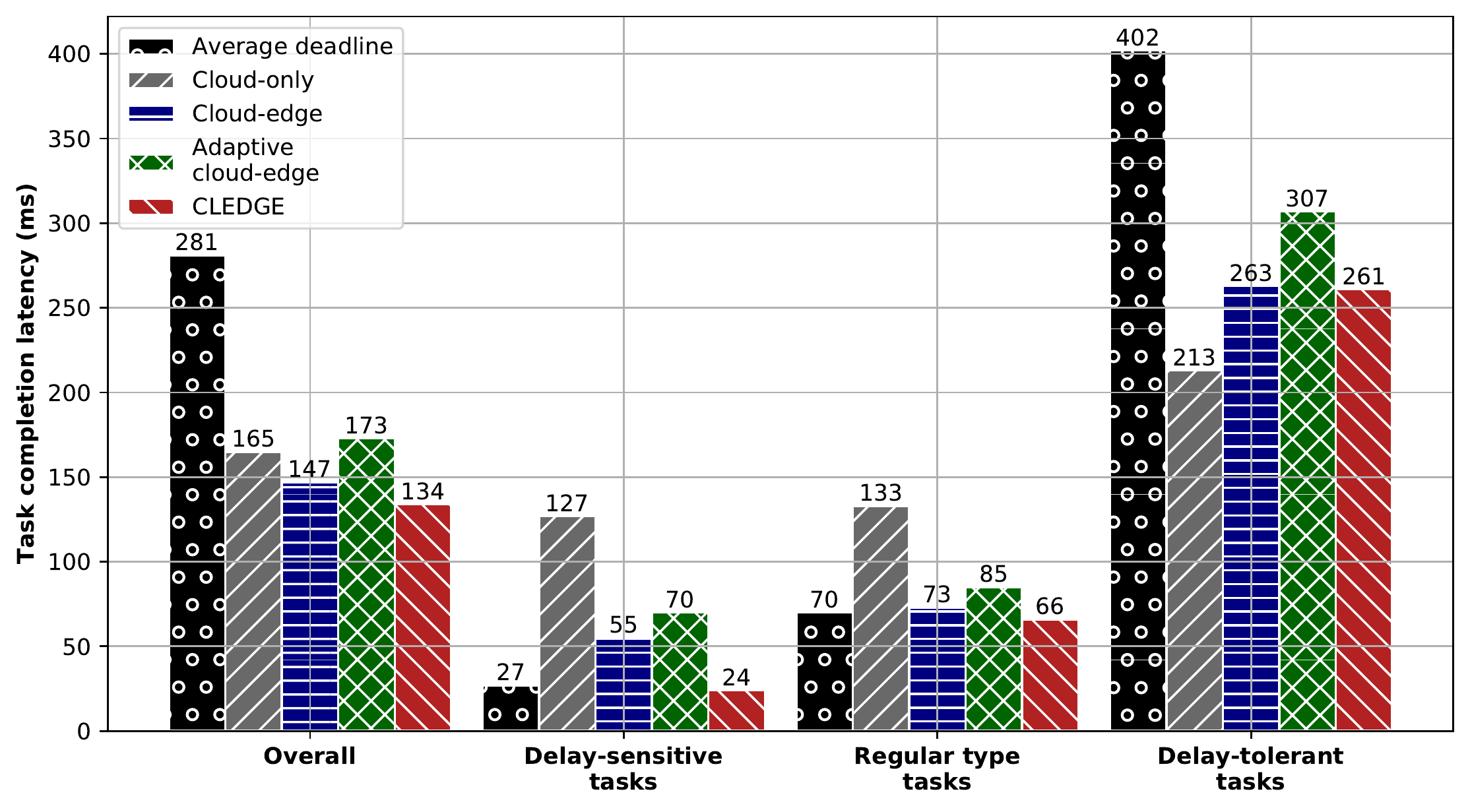}
		\subcaption{Task completion latency (heavy load)}
		\label{Figure:satisfaction-rate}
	\end{subfigure}
	\begin{subfigure}[b]{1\columnwidth}
		\centering
		\includegraphics[width=0.73\columnwidth]{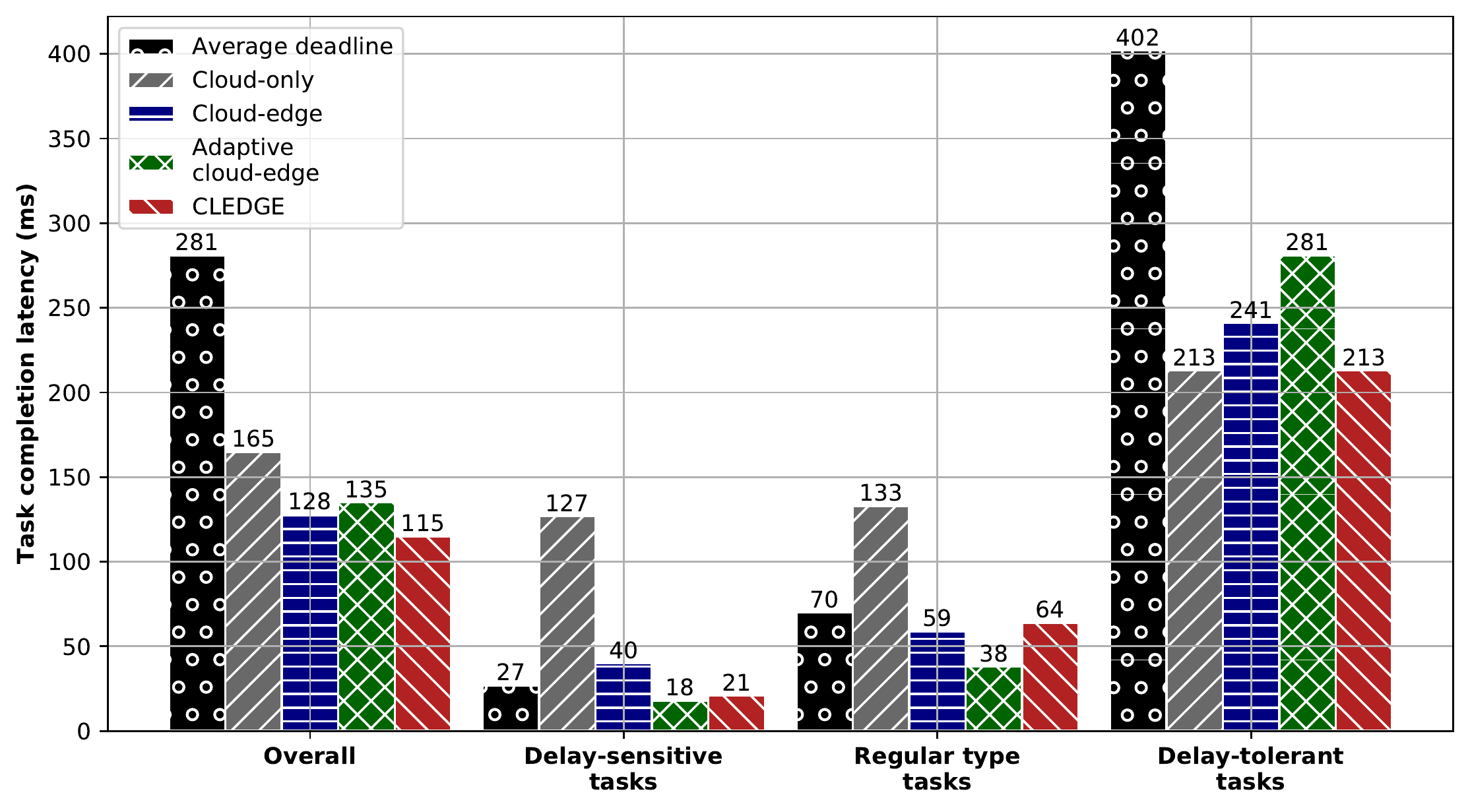}
		\subcaption{Task completion latency (light load)}
		\label{Figure:overhead}
	\end{subfigure}
	\vspace{-0.4cm}
	\caption{Average task completion latency. The results for edge-only are significantly high, thus they are omitted to improve readability. {\fgcs The ``overall'' results include the average of the results of delay-sensitive, regular, and delay-tolerant tasks.}} 
	\label{figure:latency}
\end{figure}

\noindent \textbf{Task completion latency:} In Figure~\ref{figure:latency}, we present the average completion latency for all tasks and for each service category. Our results demonstrate that \sol is the only approach that achieves completion times lower than the deadlines of tasks associated with services of all the different categories. This indicates that \sol can successfully distribute tasks for execution to available computing resources based on their deadlines, being able to satisfy categories of tasks/services with diverse latency requirements. \sol also achieves the lowest overall task completion latency (\ie average completion time among tasks of all categories) and meets the latency requirements of task categories that other approaches cannot (\eg delay-sensitive and regular type tasks under heavy loads).


\begin{figure}[th]
	\centering
	\captionsetup[subfigure]{aboveskip=-1pt,belowskip=-1.5pt}
	\begin{subfigure}[b]{0.49\columnwidth}
		\centering
		\includegraphics[width=1\columnwidth]{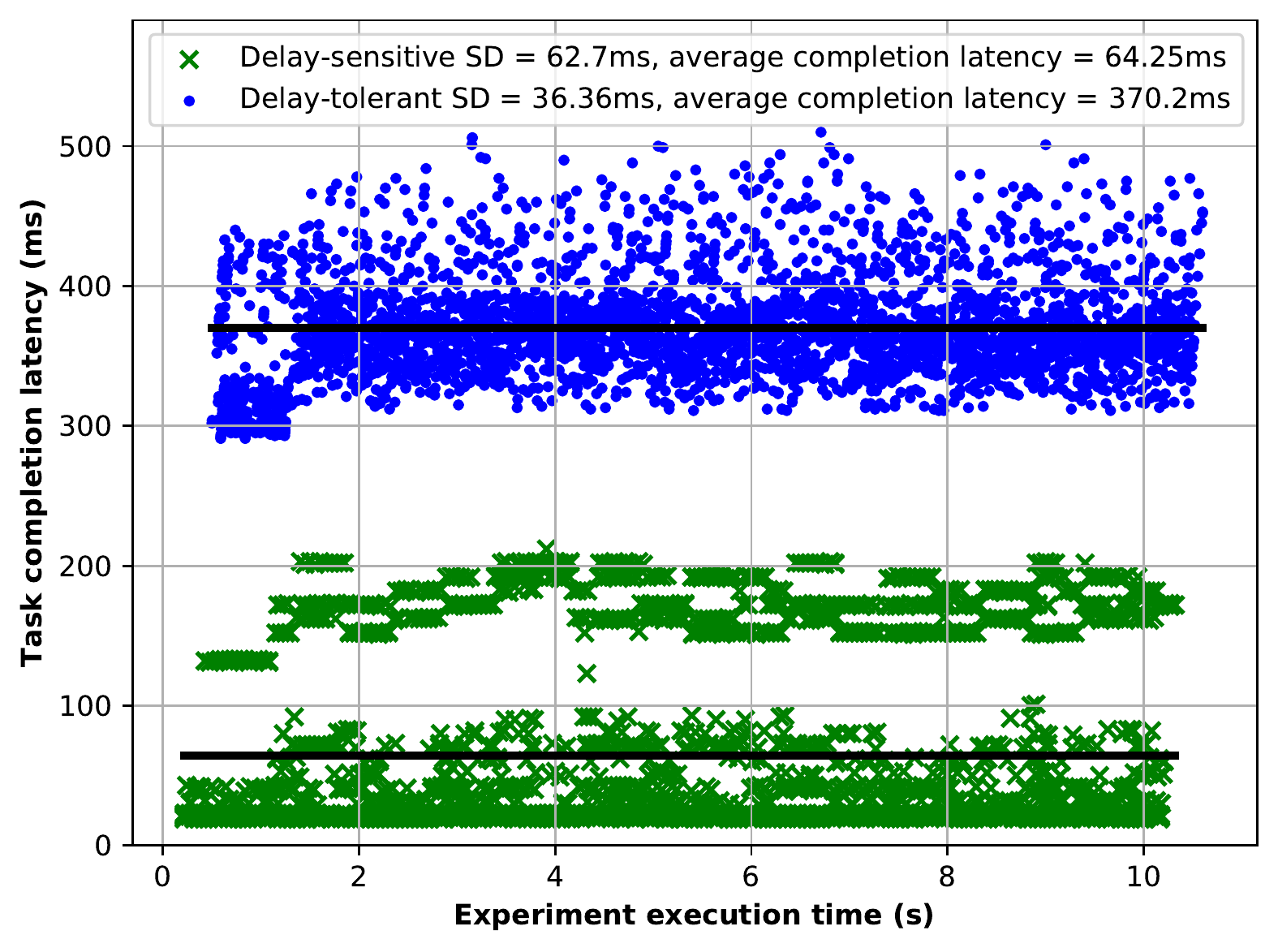}
		\subcaption{Reliability of task completion latency for adaptive cloud-edge}
		\label{Figure:alg1}
	\end{subfigure}
	\begin{subfigure}[b]{0.49\columnwidth}
		\centering
		\includegraphics[width=1\columnwidth]{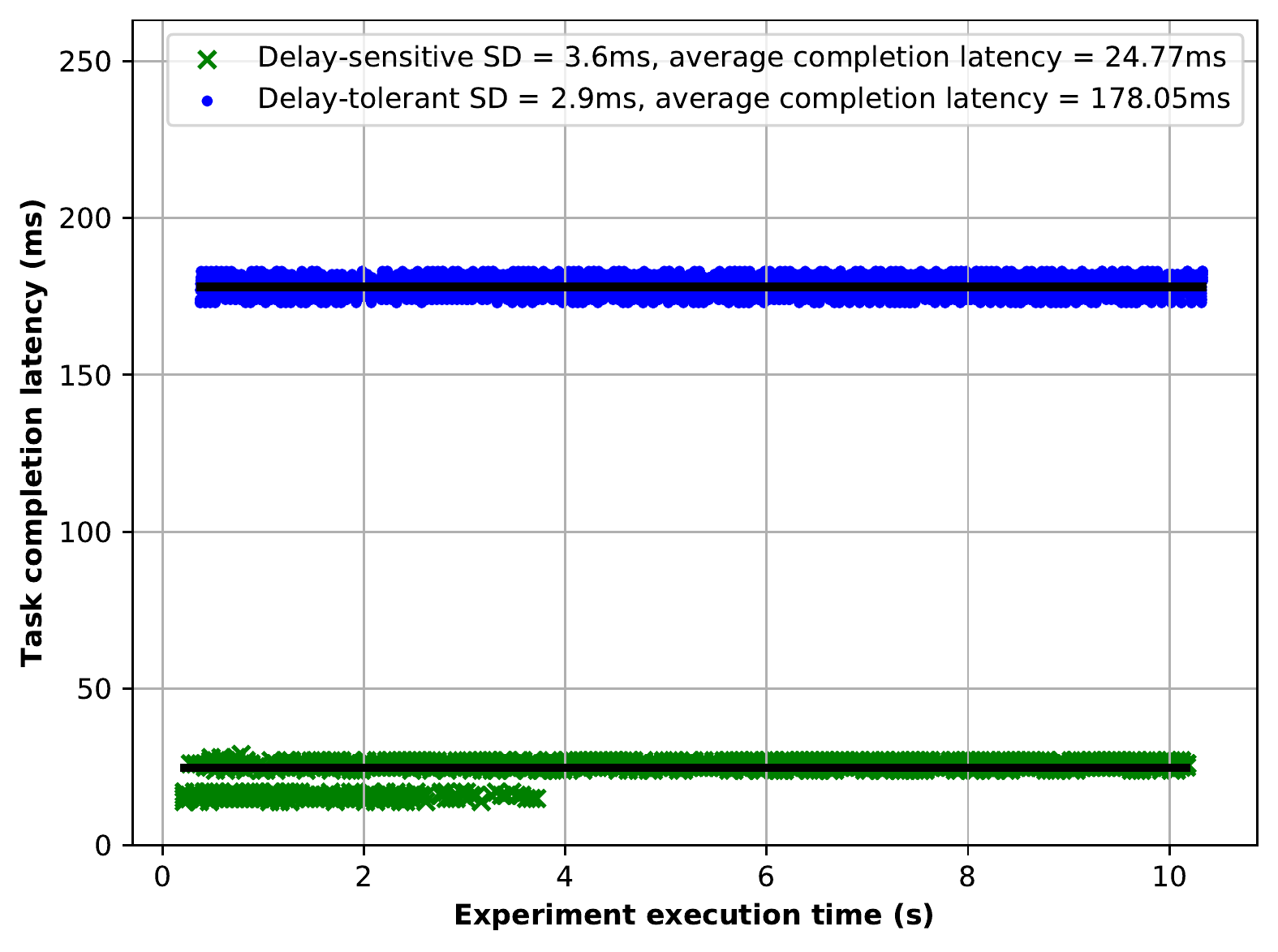}
		\subcaption{Reliability of task completion latency for \sol}
		\label{Figure:alg2}
	\end{subfigure}
	\caption{Reliability of task completion latency for a sample time interval (10 seconds) under heavy load. Results for completion latency and Standard Deviation (SD) are presented for the tasks of a delay-sensitive and delay-tolerant service.} 
	\label{figure:reliability}
	\vspace{+0.1cm}
\end{figure}

\noindent \textbf{Reliability of task completion latency:} In Figure~\ref{figure:reliability}, we show results on the reliability of the task completion latency for \sol and adaptive cloud-edge for a sample time interval of 10 seconds during our experiments under heavy load. Note that we present results for a selected delay-sensitive and delay-tolerant service. We have verified that these results are representative of the results for all other services of the same nature. Our results indicate that \sol achieves consistent latency for both delay-sensitive and delay-tolerant tasks with a minimal standard deviation of 3.6ms and 2.9ms respectively. On the other hand, adaptive cloud-edge results in inconsistent latency with a standard deviation of 62.7ms and 36.36ms for delay-sensitive and delay-tolerant tasks respectively. We verified that all other approaches result in inconsistent latency and follow the same trend as adaptive cloud-edge. 

\noindent \textbf{Distribution of executed tasks:} Through our evaluation, we were able to identify whether offloaded tasks were executed at the edge or on the cloud. We further identified whether the tasks were executed in the same or a different edge network than the one they were offloaded onto. Our results indicated that \sol executed 41-44\% of the tasks within their home edge networks (\ie the same edge networks the tasks were offloaded onto), 11-15\% across different edge networks, and 41-48\% on the cloud. This is attributed to the fact that \sol distributed delay-sensitive tasks to computing resources within their home edge networks, while tasks were distributed to neighboring edge networks only when their home edge networks did not have available resources and the tasks' deadlines did not allow for execution on the cloud. 

\begin{figure}[th]
	\centering
	\vspace{-0.35cm}
	\captionsetup[subfigure]{aboveskip=-1pt,belowskip=-1pt}
	\begin{subfigure}[b]{1\columnwidth}
		\centering
		\includegraphics[width=0.57\columnwidth]{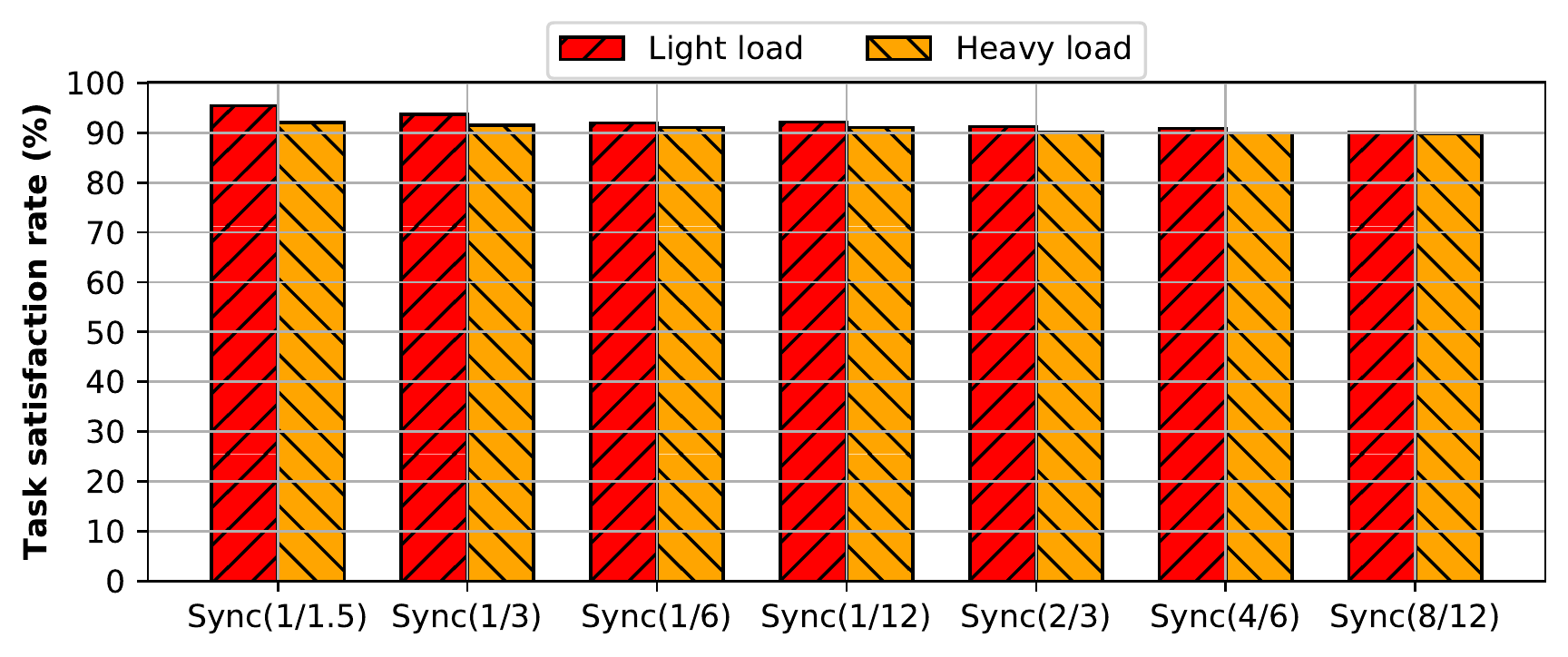}
		\subcaption{Synchronization frequency impact on \sol's satisfaction rate}
		\label{Figure:sync1}
	\end{subfigure}
	\begin{subfigure}[b]{1\columnwidth}
		\centering
		\includegraphics[width=0.57\columnwidth]{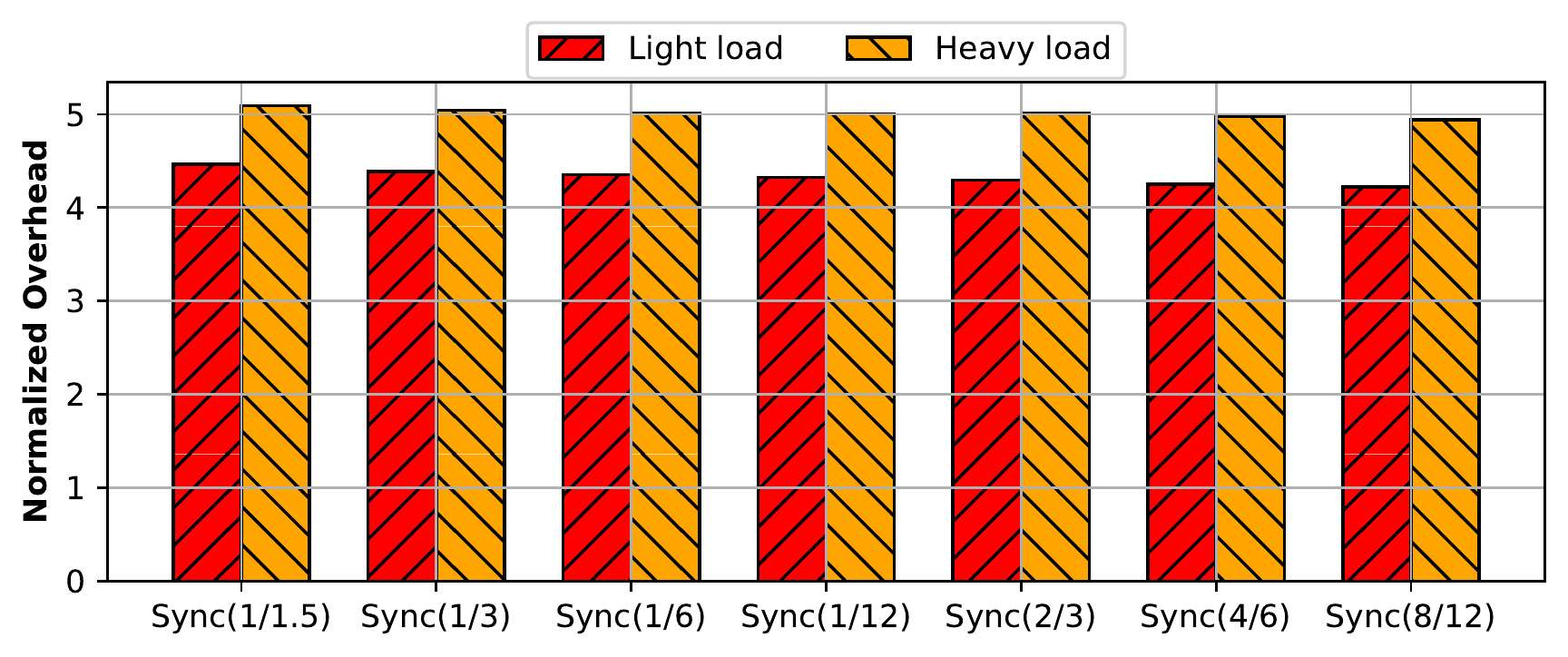}
		\subcaption{Impact of synchronization frequency on \sol's overhead}
		\label{Figure:sync2}
	\end{subfigure}
	\caption{Impact of the two-tier synchronization frequency on the task satisfaction rate and overhead. The notation $Sync(x/y)$ denotes that synchronization takes place within edge networks every $x$ seconds and among \ONs every $y$ seconds.} 
	\label{figure:syncresults}
\end{figure}

\noindent \textbf{Impact of synchronization frequency:} In Figures~\ref{Figure:sync1} and~\ref{Figure:sync2}, we present results on the impact of the frequency of the two-tier synchronization process on the task satisfaction rate and the overhead of \sol. Our results indicate that the synchronization frequency does not have a major impact on these metrics under both light and heavy loads. As synchronization becomes less frequent, the satisfaction rate and the overhead decrease by 5\%. Since the light and heavy load profiles do not include rapid task offloading rate changes, 
the results demonstrate that the synchronization period can be relatively long (in the order of several seconds). We further performed experiments where users continuously switched between light and heavy loads, signifying that the synchronization frequency can impact the satisfaction rate when rapid changes of the utilization of the resources happen. In such cases, \ENs may send an explicit notification to synchronize with others when they detect a rapid change of their resource utilization. 

\section{Conclusion and Future Work}
\label{sec:conclusion}

\vspace{-0.1cm}

In this paper, we presented \sol, an NDN framework for hybrid cloud-edge computing. Its design was motivated by the requirements of the mixed reality community for the execution of tasks with diverse deadlines along with reliable response times. Our evaluation indicated that \sol can provide adaptive distribution of tasks based on the latency they can tolerate to available computing resources within the same or different edge networks, and towards the cloud. 

In our future work, we will investigate the following directions: (i) sophisticated synchronization mechanisms that provide beneficial trade-offs between task satisfaction rates and overheads; (ii) utilize \sol to facilitate the operation of a wide range of applications; 
and (iii) implement a \sol prototype, conduct a real-world evaluation study of its design, performance, and scalability.



\vspace{-0.2cm}


\section*{Acknowledgements}

This work is partially supported by the National Science Foundation under awards CNS-2016714, CNS-2104700, OAC-2019163, and OAC-2019012, the National Institutes of Health (NIGMS/P20GM109090), the Nebraska University Collaboration Initiative, and the Nebraska Tobacco Settlement Biomedical Research Development Funds.

\balance
\bibliographystyle{unsrt}
\bibliography{bib}


\end{document}